\documentclass[preprint2]{aastex}

\slugcomment{Submitted to The Astrophysical Journal }

\shorttitle{X-ray Variability of AGNs }
\shortauthors{Chitnis et al.}

\begin{document}

\title{X-ray variability of AGNs in the soft and the hard X-ray  bands}

\author{V. R. Chitnis}
\affil{Department of High Energy Physics, Tata Institute of Fundamental Research, 
Mumbai 400005, India }

\author{J. K. Pendharkar}
\affil{Indian Institute of Astrophysics, II Block, Koramangala, Bangalore 560 034, 
India, and Physical Research Laboratory, Ahmedabad - 380 009 }

\author{D. Bose\altaffilmark{1}}
\affil{Department of Astronomy and Astrophysics, Tata Institute of Fundamental Research,
Mumbai 400005, India }

\author{V. K. Agrawal}
\affil{Department of Astronomy and Astrophysics, Tata Institute of Fundamental Research,
Mumbai 400005, India}

\author{A. R. Rao}
\affil{Department of Astronomy and Astrophysics, Tata Institute of Fundamental Research,
Mumbai 400005, India }

\author{R. Misra}
\affil{Inter-University Center for Astronomy and Astrophysics, Post Bag 4,
Ganeshkhind, Pune-411007, India}

\altaffiltext{1}{present address : Universidad Complutense, E-28040 Madrid, Spain}

\begin{abstract}
 We investigate the X-ray variability characteristics of hard X-ray
selected  AGNs  (based on Swift/BAT data) in the soft X-ray band 
using the RXTE/ASM data. The uncertainties  involved in the individual 
dwell measurements of ASM are critically examined  and a method is
developed to
combine a large number of dwells with appropriate error 
propagation to derive long duration flux measurements (greater than 10 days).
We also provide a general prescription to 
estimate the errors in variability derived from rms values 
from unequally spaced data. Though the derived variability for
individual sources are not of very high significance, we find that, in general,
the soft X-ray variability is higher than those in hard X-rays and
the variability strengths decrease with energy for the  diverse classes
of AGN.
We also examine the
strength of variability as a function of the break time scale in
the power density spectrum (derived from the estimated mass and
bolometric luminosity of the sources) and find that 
the data are consistent with the idea of higher variability at time
scales longer than the break time scale.

\end{abstract}


\section{Introduction}

Time variable X-ray emission is a probe to the inner regions of the Active Galactic Nuclei 
(AGN) and is considered to be one of their defining characteristics. 
X-ray 
variability studies provide 
insight into the geometry and the physical conditions in the nuclear regions on account 
of the fact that this emission is thought to be emitted from regions close to  the 
supermassive black hole. 
The X-ray  variability studies
on time-scales of months to years in AGNs have established 
the similarity of the physical processes across stellar mass to supermassive black holes \citep{utt04,mch06}.
Since the characteristic time-scale is proportional to the mass of 
the black hole, $M_{BH}$, continuous monitoring for a long duration becomes necessary in 
the case of AGN  based on linear scaling with black hole mass from X-ray binary systems.

Early studies on X-ray variability using data from {\em EXOSAT} in the 0.1 -- 10 keV range 
showed that on short time-scales AGN variability appeared to be red noise
dominated. In other
words, it was unpredictable and aperiodic in nature \citep{mch87}. Corresponding
power spectral density (PSD, that is, variability power as a function of temporal frequency)
is best fitted by  a  power-law of slopes -1 to -2 with no cut-off seen down to the lowest 
sampled frequencies.
However, the shapes of their PSDs were shown to be similar to those 
of X-ray binaries (XRBs) in their soft state 
\citep{mch88}.
Hence analogous to the 
high-frequency breaks seen in the PSDs of XRBs and assuming the break time-scale varying 
linearly with $M_{BH}$, a break to flatter PSD slope was expected over a period of days 
to weeks. Due to the uneven sampling of data the results yielding the break-frequency were 
uncertain. The {\em EXOSAT} data of a $\sim$day-long AGN X-ray observations have revealed
inverse correlation between the amplitude of variability and X-ray luminosity \citep{bar86}. 
Results from the {\em ROSAT} All-Sky Survey (RASS) on the soft X-ray
variability of AGN showed the variability strength on timescales of days to be a function
of steepness of the X-ray spectrum with sources with steeper spectra exhibiting stronger 
variability \citep{gru01}. PSD studies were constrained due to large time gaps 
between subsequent observations in {\em ROSAT} data.

With the launch of the {\em RXTE} in 1995, a significant improvement was seen in the quality 
of data and a manifold increase in the monitoring timescale, from months to years, was 
possible. Because 
of its rapid slewing capability and flexible scheduling, an evenly sampled long-term 
monitoring of AGN X-ray variability was carried out for several sources. 
Contrary to the results from the then existing satellites, AGN light curves from
{\em RXTE/PCA} in the 2 -- 20 keV energy band showed that on longer timescales, of
about a month, sources displayed less dispersion in variability amplitudes compared
to those measured on time scales of 1 day \citep{mar01}.
The PSDs also revealed 
a cutoff/ break at long timescales correlated with their black hole mass \citep{ede99,utt02,mar03}. 
but with a large scatter in the correlation \citep{don05}.
\citet{mch06} ascribed this scatter to a third variable,
accretion rate, and found that the break time scale combined with accretion rate 
can predict the mass of the black hole all the way from Galactic X-ray binaries
to supermassive AGNs. They postulated that the variability originates within
the accretion disk and the break time scale is associated with the inner edge of 
the disk, which makes an inward movement with an increasing accretion rate
for a given black hole mass.

 Since such a scenario makes a definitive suggestion of the accretion disk geometry,
a study of variability with energy should be able to pin down the radiation
processes in the inner accretion disk quite reliably.
Recently, the variability studies of AGN at energies above 15 keV are carried
out by \citet{bec07}. 
The data of the first 9 months of the {\em Swift/BAT} all-sky 
survey in the 14 - 195 keV range of the 44 brightest AGN revealed a tendency of
unabsorbed or type 1 Seyfert galaxies to be less variable than absorbed or type 2 
objects. Also they found a more solid anti-correlation between variability and luminosity,
which was previously detected in soft X-rays, UV and optical bands.

The RXTE/ASM all-sky survey data gives information on a large number of X-ray sources,
including AGNs. 
The sensitivity (typically 10 mCrab for the one day average data)
is not sufficient to make a
detailed study of AGNs. By taking data at larger bin sizes, however, it
should be possible to get meaningful information on the variability at 
longer time scales. 
In this paper, we have carried out a comparative study of the variability of AGN in
soft ($<$ 12 keV) and hard ($>$ 12 keV) X-ray bands using data from the {\em RXTE/ASM} 
all-sky survey and
results obtained by Beckmann et al. from {\em Swift/BAT} all-sky survey. Data 
selection is described in section 2 followed by data analysis and results in section 3 
and discussion and conclusions in section 4.

\section{Data Selection} 

The All-Sky Monitor (ASM, \citet{lev96}) onboard  Rossi X-ray Timing Explorer 
({\em RXTE}, \citet{swa99}) consists of three Shadow Scanning Cameras 
(SSCs). Each SSC contains a position-sensitive proportional counter (PSPC) that views
sky through a slit mask. FOV of each SSC is 6$^\circ$$\times$90$^\circ$, allowing ASM to scan
most of the sky every 1.5 hours. So apart from locating transient objects, it also 
provides 
photometric records of known sources in three energy bands corresponding to A 
(1.5 -- 3 keV), B (3 -- 5 keV) and C (5 -- 12 keV) in addition to the total, or sum band, 
intensity in the 1.5 -- 12 keV band. 
The MIT database (http://xte.mit.edu/ASM\_lc.html), 
from where we have extracted ASM 
light curves, gives data dwell by
dwell and one day average data points. In dwell by dwell each raw data point represents
the fitted source flux from one 90 second dwell. Whereas in case of 'one day average'
each data point represents the one-day average of the fitted source fluxes from a number 
(typically 5 -- 10) of individual ASM dwells. 

To have a comprehensive information, we have selected all AGN candidates (147) from
the ASM source list and carried out a systematic analysis. We found 31 AGN from the 
sample used by \citet{bec07}. We find measurable ASM flux for all
the sources except two (NGC 1365 and GRS 1734-292).
  This sample of common sources consists of 3 blazars,
1 radio galaxy, 8 Seyfert 1, 4 Seyfert 1.5, 2 Seyfert 1.8, 1 Seyfert 1.9 and 12
Seyfert 2. The BAT sources not in the ASM 
source list include  1 blazar, 3 Seyfert 1, 
1 Seyfert 1.5 and 8 Seyfert 2. 


\section{Data Analysis and Results}

\subsection{ASM dwell data selection criteria} 

\begin{figure}
\epsscale{1.0}
\plotone{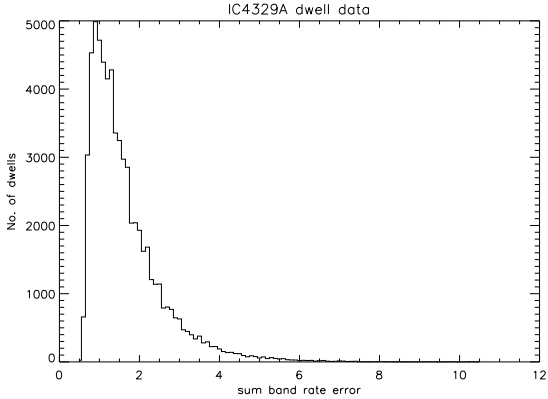}
\caption{Distribution of errors on sum band rates from dwells for IC4329A
\label{dwellerrdist}}
\end{figure}

We have downloaded all available ASM dwell data for 147 AGN from MIT database 
(http://xte. mit.edu/ASM\_lc.html). This covers data from MJD 50087 to about
54466, i.e. 1996 January 5 to 2008 July 15. Dwell data are 
subjected to the usual  selection cuts  as prescribed by the ASM website and 
binned in 20 and 40 days bins. 
On closer inspection of light curves of individual sources, several large
spikes, some of them with an apparent periodicity of one year, were noticed.
These abnormal data were found to have large intrinsic errors too. Since
the count rates in ASM is based on a profile fitting method, the errors 
depend on the other sources in the field of view and the orientation of the
ASM FOV for a given observation. As a first cut, we examined the possibility that
there could be enhanced unaccounted systematic errors for data points 
with large measured errors. It is found that the measured errors for each
dwell data is sharply peaked at 1 counts/s.  The distribution of errors
for each dwell data for one of the sources (IC4329A) is shown in 
Fig.~\ref{dwellerrdist}. Based on these considerations, we have introduced
a selection criterion 
of an upper limit on sum count rate error of 3 counts/s, 
corresponding  to thrice the peak value of the distribution.

\begin{figure}
\epsscale{1.0}
\plotone{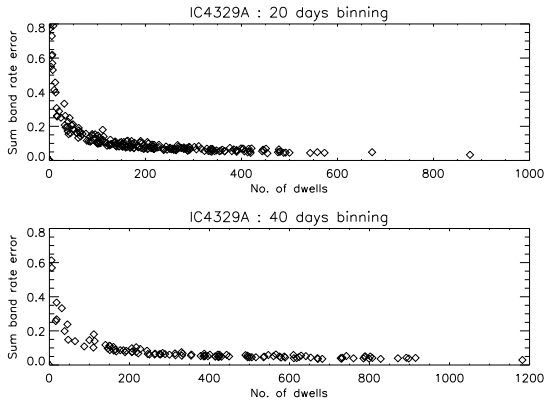}
\caption{Sum rate error vs no. of dwells for IC4329A with 20 days and
40 days binning.  \label{errorvsdwell}}
\end{figure}

After selecting dwell count rates for sum, A, B and C bands based on the  above
mentioned cuts, they are binned into 20 and 40 days starting from MJD
50087. 
 For each bin the average rate and error are calculated for all the 
four bands, which are weighted mean and estimated error of the mean.
 Data points with too few number of individual dwell measurements
(less than 20 data points for 20 day binning and less than 40 data points for 
40 day binning) are also ignored, ensuring that there is at least one dwell per
day, on an average.
For example, for  IC4329A, Fig.~\ref{errorvsdwell}
shows the  sum band rate error after binning as a function of number of dwells per bin,
for 20 and 40 days binning. In both the cases sum band error increases
sharply for bins with lower number  of dwells. 
After introducing these criteria, the individual light curves are carefully
examined and no systematic abnormalities are found in these light curves.

 To summarize, 
the selection criteria for generating ASM light curves
are:

\begin{itemize}
  \item [1.]  Reduced chi-square of the fit $<$ 1.5
  \item [2.] Number of sources in the field of view $<$ 16
  \item [3.] Earth angle $>$ 75$^\circ$
  \item [4.] Exposure time $>$ 30 seconds
  \item [5.] Long-axis angle -41.5$^\circ$ $<$ $\theta$ $<$ 46$^\circ$
  \item [6.] Short-axis angle -5$^\circ$ $<$ $\phi$ $<$ 5$^\circ$
  \item [7.] Error on sum count rate $<$ 3 counts/s
  \item [8.] Number of dwells for N day bin light curve is  $>$ N
\end{itemize}

The first six selection criteria are the same as those used for generating one-day
average count rates from dwell data as described in the  MIT database webpage.

\subsection{ASM light curves}

The ASM measurements have a systematic error of 3\% added in quadrature to
the statistical errors. This error is estimated using Crab light curves and
could be underestimate in some cases (http://xte.mit.edu/ASM\_lc.html). 
\citet{gri02} have studied light curves of several sources with constant
X-ray flux binned with different bin durations from 1 to 200 days and have 
estimated systematic error on large time scales in the range of 0.01-0.1
cts/s, depending on the source flux. According to Fig. 2 of their paper, there
seems to be a shift of about 10\% between measured and estimated RMS for 
sources with constant flux. We have investigated this in detail to get a 
better handle on the average systematic errors. For a few of the
bright sources, we have generated the  power spectral density spectra (PSD)
using a method based on the autocorrelation function (Gilfanov \& 
Arefiev 2005)
and have assumed that the frequency independent power has to be 
due to the errors in each observations.  Based on these PSD, rms noise 
level seems to be underestimated by a factor of 1.13.
A few steady sources too were investigated and the frequency independent
systematic error is similar to this number. Though we cannot completely
rule out some additional frequency dependent
systematic errors, we assume that for large bin sizes (greater than 10 days),
most of the systematic errors are taken into account. This assumption
is further strengthened by examining individual light curves as well as
the distribution of the source fluxes (see below).
After correcting typical binned error of 0.005 c/s with this factor, 
this systematic error translates into typical systematic error of 0.5 c/s
for dwell in S band and 0.3 c/s for A, B and C bands.

\begin{figure}
\epsscale{.80}
\includegraphics[scale=.40,angle=270]{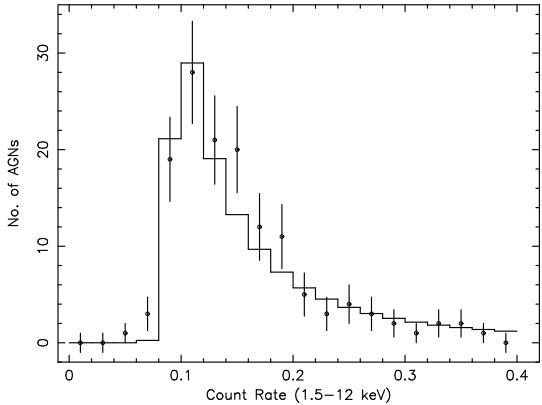}
\caption{Histogram of ASM Sum band count rates for 147 AGN. Data is shown
as points with error bars and the
fit by power law with constant offset is shown as a histogram.
\label{sumratehist}}
\end{figure}

According to ASM data products page 
(http:// xte.mit.edu/), there is a
1 mCrab positive bias in the light curve intensities. This bias is only 
evident when enough data are averaged so that the statistical uncertainties 
are driven to be very small. This generally requires binning or analyzing the 
light curves on time scales of many days or longer. In order to estimate
this bias we have carried out the following exercise. We have generated histogram
of average sum band count rates for 147 AGN (Fig.~\ref{sumratehist}). 
Unlike the usual number distribution
given in an integral way, this is a differential plot and hence the
number should vary with count rates as a power-law with an index of -2.5
for an isotropic distribution. Hence we have fitted this histogram with a
function consisting of a power-law with slope -2.5 and constant offset
represented as the start point of the power-law. 
Count rates were spread in the X-axis using a Gaussian distribution with $\sigma$=0.0056 
counts/s according to the above mentioned error estimation. Fit for sum band data
is shown in the Fig.~\ref{sumratehist}. The 
number distribution is shown as data points with error bars
(error taken is the square root of the number in each bin)
and the fit is shown as a histogram.
Offsets obtained with
this method for various bands are as follows: S band : 0.0895 c/s,
A band : 0.0274 c/s, B band : 0.015 c/s and C band : 0.0213 c/s. These
values are consistent with the 1 mCrab offset quoted in the ASM website.

\begin{figure}
\epsscale{1.0}
\plotone{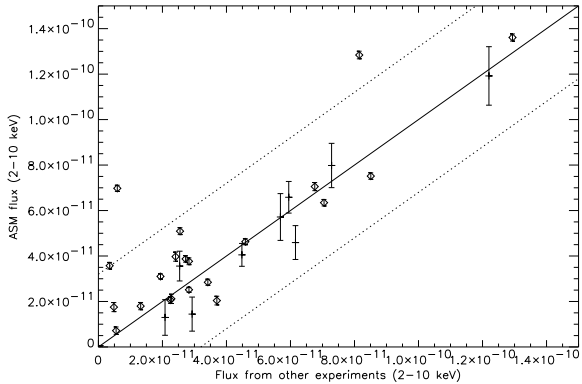}
\caption{Absorbed ASM flux vs observed flux (in units of ergs cm$^{-2}$ s$^{-1}$) from various experiments shown
by diamonds.  ASM flux is derived from average of ASM count rate over the period
of more than ten years. Solid line corresponds to to the case where ASM flux
is the same as flux from other measurements. Dotted lines correspond to ASM
flux in the range of $\pm$0.1 count/s around solid line. Points indicated by
plus signs (and large error bars)  correspond to flux measurements from PCA onboard RXTE compared
with simultaneous ASM measurements for  IC4329A, NGC4151,
NGC3783, MCG-6-30-15, 3C120, 3C390.3, NGC3227, NGC4051 and
NGC3516.
\label{asmvsxrtflux}}
\end{figure}

As a further verification of these offsets and systematic error, we checked
for consistency of flux estimates. 
We compared our results with the flux obtained from 
 an uniform spectral analysis done on a large sample of AGNs
\citep{win08} and  found 22 common sources.
Using the spectral parameters from \citet{win08} and ASM sum band average count rates we estimated 
2--10 keV absorbed flux for these sources using web calculator (WebPIMMS tool
at http://heasarc.gsfc.nasa.gov/Tools). Fig.~\ref{asmvsxrtflux} shows the
plot of  ten years averaged absorbed ASM flux vs observed flux from other experiments, 
both in the energy range of 2--10 keV. 
It should be noted that these observations are not simultaneous. Hence
source variability can cause non-identical fluxes between the ASM and the 
other experiments.
Points are marked by diamonds. Solid line corresponds 
to the case where ASM flux is equal to flux measurements from other experiments.
Dotted lines correspond to ASM flux equivalent to $\pm$0.1 c/s around solid line. 
Most of the sources barring a few like IC4329A and NGC 4388 are close to the fitted line, 
within $\pm$0.1 ASM c/s. 
As a further check we compared the long  duration  RXTE PCA measurements from the uniform 
analysis done by \citet{mar04}.
We found these measurements for nine sources: IC4329A, NGC4151, 
NGC3783, MCG-6-30-15, 3C120, 3C390.3, NGC3227, NGC4051 and 
NGC3516 \citep{mar04}. We estimated the ASM flux 
by extracting 40 days binned sum band rates near simultaneous
with the PCA measurements. These points are indicated by plus signs with error
bars in the plot. Hence overall there is a good agreement between flux estimates 
from ASM and other observations. This agreement also justifies our estimates of count rate offsets
and systematic errors.
We note here that further refinement in the flux comparison is difficult
to achieve because of the fact that at shorter time scales (where pointed
observations are available) ASM error bars are too large and at longer time
scales (when the ASM error bars are driven to a small value) 
pointed observations are simply not available. 

\begin{figure}
\epsscale{.80}
\includegraphics[scale=.32,angle=270]{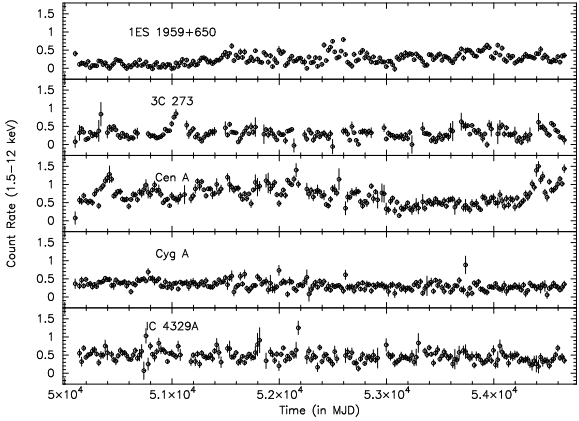}
\caption{ASM Sum band light curves (1.5-20 keV) for 1ES 1959+650, 3C 273,
Cen A, Cyg A and IC 4329A with 20 days binning. \label{sumfig1}}
\end{figure}

\begin{figure}
\epsscale{.80}
\includegraphics[scale=.32,angle=270]{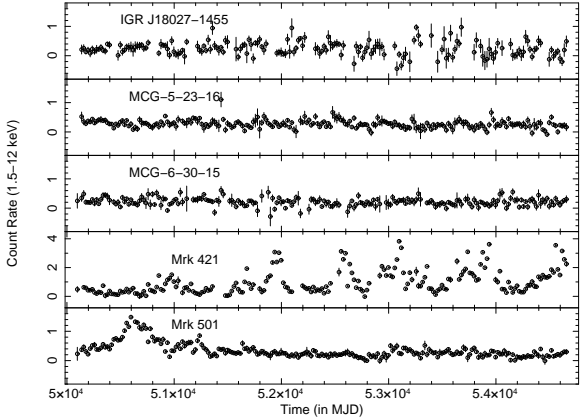}
\caption{ASM Sum band light curves (1.5-20 keV) for IGR J18027-1455,
MCG-5-23-16, MCG-6-30-15, Mrk 412 and Mrk 501 with 20 days binning.
\label{sumfig2}}
\end{figure}

\begin{figure}
\epsscale{.80}
\includegraphics[scale=.32,angle=270]{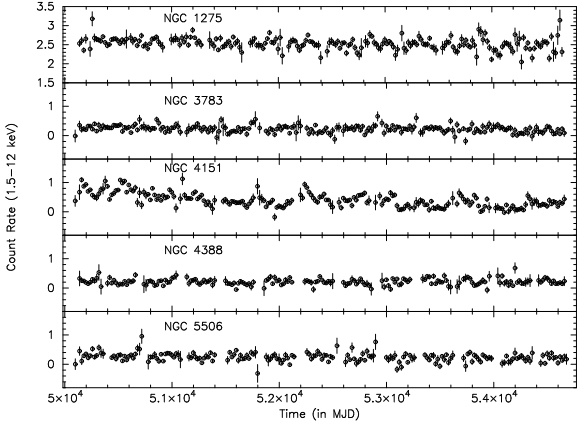}
\caption{ASM Sum band light curves (1.5-20 keV) for NGC 1275, NGC 3783,
NGC 4151, NGC 4388 and NGC 5506 with 20 days binning. \label{sumfig3}}
\end{figure}

Applying the above mentioned cuts and corrections, light curves in all four bands 
were generated for all ASM detected AGNs. Average sum band count rates of all these
AGN in four bands with 20 days binning for 30 sources common to ASM and BAT 
database are given in Table~\ref{tab_var20}. There was one more common
source, GRS 1734-292, which was rejected after applying above the mentioned cuts. 
That is, there was not a single stretch of 20 days with at least 20 dwells
satisfying all the cuts (possibly because of its proximity to the Galactic center).
Count rates given in Table~\ref{tab_var20} are weighted mean of binned counts rates 
and estimated error on mean. 
Out of these, 15 sources have average sum band count rates above 0.2 counts/s. 
Some of these sources (11) are common sources in BAT and ASM database and are included
in Table ~\ref{tab_var20}. The remaining sources are 1ES1959+650, IGR J18027-1455,
MCG 6-30-15 and Mrk 501.
Light curves for these 15 sources with 20 days binning are given in 
Figs.~\ref{sumfig1}, ~\ref{sumfig2} and ~\ref{sumfig3}.

\begin{deluxetable}{cccccccccccc}
\tabletypesize{\scriptsize}
\rotate
\tablecaption{Variability strength and T$_{Break}$ for ASM sample \label{tab_var20}}
\tablewidth{0pt}
\tablehead{
\colhead{Source} & \colhead{ASM rate} & \colhead{ASM } & \multicolumn{3}{c} {ASM variability strength\tablenotemark{1}} &
\colhead{BAT\tablenotemark{1}} & \colhead{BH Mass} & \colhead{log(L$_{bol}$)} &
\colhead{log($T_B$)} & \colhead{Type}
 \\
\colhead{name} & \colhead{(1.2-15 keV)} & \colhead{luminosity} & \multicolumn{3}{c}{S$_v$} &
\colhead{variability} & \colhead{(log M$_{BH}$)} &
\colhead{ergs s$^{-1}$} & \colhead{days} & \colhead{} \\
\colhead{} & \colhead{counts s$^{-1}$} & \colhead{log(L$_{ASM}$)} & \multicolumn{3}{c}{} &
\colhead{strength} & \colhead{M${sun}$} & \colhead{} &
\colhead{} & \colhead{} \\
\colhead{} & \colhead{} & \colhead{ergs s$^{-1}$} & \colhead{1.5-12 keV} & \colhead{1.5-3 keV} & \colhead{3-12 keV}  &
\colhead{S$_v$} & \colhead{} & \colhead{} & \colhead{} & \colhead{}
}
\startdata
3C273           &0.280$\pm$0.004  &45.68  &   34.7$\pm$3.4   & 64.7$\pm$7.9   & 33.6$\pm$4.4   & 15$\pm$5   &  7.22  &47.35    &-3.04&   Blazar\\
3C454.3         &0.078$\pm$0.006  &46.59  &     ...          &     ...        &     ...        &42$\pm$12   &  9.17  &47.27    &1.13 &  Blazar\\
Mrk421          &0.894$\pm$0.005  &44.57  &   83.7$\pm$4.2   & 79.4$\pm$4.1   & 88.4$\pm$4.5   &142$\pm$38  &  8.29  &$\sim$45.00  &1.51  &  Blazar\\
IGR J21247+5058 &0.126$\pm$0.004  &43.54  &   42.3$\pm$5.0   &     ...        & 51.4$\pm$5.0   & 11$\pm$6   &  ...   & ...      &  ...   & Radio galaxy\\
3C120           &0.173$\pm$0.005  &44.11  &   42.7$\pm$6.3   & 64.1$\pm$7.9   & 38.9$\pm$10.4  &   $\le$10  &  7.42  &45.34     &-0.65   & Sy1\\
3C390.3         &0.109$\pm$0.004  &44.37  &   29.7$\pm$5.3   & 49.6$\pm$9.4   & 30.5$\pm$7.5   &   $\le$6   &  8.55  &44.88     &2.17    & Sy1\\
EXO0556-386     &0.069$\pm$0.005  &43.74  &       ...        & 76.8$\pm$14.1  &     ...        &     $\le$9  &  ...   & ...      & ...    & Sy1\\
IC4329A         &0.459$\pm$0.006  &43.91  &   22.7$\pm$2.5   & 48.0$\pm$4.5   & 18.1$\pm$3.5   & $\le$3     &  6.77  &44.78    &-1.47&   Sy1\\
MR2251-178      &0.134$\pm$0.005  &44.57  &   51.2$\pm$8.4   & 96.7$\pm$22.5  & 42.3$\pm$13.5  &   $\le$7   &  ...   & ...      & ...    & Sy1\\
NGC3783         &0.222$\pm$0.005  &43.16  &   28.5$\pm$4.9   & 67.3$\pm$12.8  & 24.1$\pm$7.4   &   $\le$4   &  6.94  &44.41    &-0.75 &   Sy1\\
NGC4593         &0.105$\pm$0.005  &42.76  &   79.7$\pm$11.2  &     ...        & 71.3$\pm$21.8  &   $\le$7   &  6.91  &44.09     &-0.50   & Sy1\\
NGC3227         &0.125$\pm$0.005  &42.10  &   56.2$\pm$7.1   & 123.2$\pm$17.6 & 49.3$\pm$11.4  &  $\le$14   &  7.64  &43.86    &1.26 &  Sy1.5\\
NGC3516         &0.091$\pm$0.004  &42.68  &   76.6$\pm$7.0   &     ...        & 72.0$\pm$9.6   &   $\le$7   &  7.36  &44.29     &0.25    & Sy1.5\\
NGC4051         &0.064$\pm$0.005  &41.31  &       ...        &     ...        &     ...        &    $\le$9  &  6.13  &43.56     &-1.62   & Sy1.5\\
NGC4151         &0.434$\pm$0.005  &42.51  &   53.9$\pm$3.2   & 136.3$\pm$12.5 & 53.3$\pm$3.2   & 27$\pm$7  &  7.13  &43.73    &0.32 &  Sy1.5\\
NGC1365         & $<$0.005        &...    &     ...          & ...            &  ...           &  $\le$17  &  ...   & ...     & ... &  Sy1.8\\
MCG-05-23-016   &0.263$\pm$0.005  &43.11  &   32.8$\pm$3.6   &     ...        & 32.2$\pm$4.6   & 6$\pm$4   &  7.60  &44.21    &2.85 &   Sy1.9\\
NGC5506         &0.237$\pm$0.006  &42.79  &   42.9$\pm$4.6   &     ...        & 47.7$\pm$5.2   &   $\le$3  &  8.00  &44.53    &1.36 &   Sy1.9\\
CenA            &0.674$\pm$0.006  &42.19  &   32.7$\pm$1.9   & 73.4$\pm$7.9   & 35.9$\pm$2.0   & 10$\pm$2   &  8.38  &43.0     &3.66 &   Sy2\\
CygnA           &0.331$\pm$0.005  &44.85  &   25.8$\pm$2.4   & 36.0$\pm$4.1   & 27.7$\pm$3.0   &  $\le$7   &  9.40  &46.0     &2.86 &  Sy2\\
ESO103-G35      &0.120$\pm$0.006  &43.16  &   69.5$\pm$9.0   & 76.8$\pm$14.1  & 74.3$\pm$10.1  &  $\le$8   &  ...   &...      &...  &   Sy2\\
Mrk348          &0.054$\pm$0.006  &42.91  &       ...        &     ...        &     ...        & 12$\pm$10 &  7.21  &44.27    &-0.04&   Sy2\\
NGC1275         &2.541$\pm$0.005  &44.73  &   4.62$\pm$0.41  & 5.6$\pm$0.8    & 4.7$\pm$0.5    &  $\le$14  &  8.51  &45.04    &1.93 &  Sy2\\
NGC2110         &0.168$\pm$0.005  &42.84  &   41.7$\pm$5.3   & 69.7$\pm$13.9  & 28.4$\pm$7.6   & 25$\pm$7   &  8.30  &44.10    &2.41 &  Sy2\\
NGC2992         &0.122$\pm$0.005  &42.69  &  101.7$\pm$8.1   & 128.1$\pm$14.4 & 118.6$\pm$11.1 & 45$\pm$19  &  7.72  &43.92    &1.37 &   Sy2\\
NGC3081         &0.022$\pm$0.005  &41.98  &       ...        &     ...        &     ...        & 23$\pm$11  &  ...   & ...     & ... &   Sy2\\
NGC4388         &0.204$\pm$0.004  &42.99  &   17.6$\pm$7.8   & 36.0$\pm$10.6  & 32.6$\pm$15.3  & 11$\pm$4   &  ...   & ...     & ... &   Sy2\\
NGC4507         &0.062$\pm$0.006  &42.77  &      ...         &     ...        &     ...        &  $\le$8   &  ...   & ...     & ... &   Sy2\\
NGC7172         &0.063$\pm$0.006  &42.52  &      ...         & ...            &     ...        &12$\pm$9   &  ...   & ...     & ... &   Sy2\\
NGC7582         &0.056$\pm$0.005  &42.02  &      ...         &     ...        &     ...        &23$\pm$21  &  ...   & ...     & ... &   Sy2\\
Mrk3            & ...             &...    & ...              &...             & ...            &   $\pm$7   &  8.65  &44.54    &2.72 &   Sy2\\
1ES1959+650     &0.243$\pm$0.003  &44.26  &   55.8$\pm$3.0   & 56.9$\pm$3.4   & 60.4$\pm$3.6   & ...       &         &         &     & Blazar\\
IGRJ18027-1455  &0.254$\pm$0.010  &43.64  &   42.3$\pm$5.0   & 77.1$\pm$14.7  & 70.4$\pm$12.1  & ...       &         &         &     &   Sy1\\
MCG-6-30-15     &0.208$\pm$0.006  &44.65  &   34.3$\pm$5.7   & 69.1$\pm$7.8   & 40.2$\pm$8.7   & ...       &  6.65   & 43.56   & -0.52 & Sy1\\
Mrk 501         &0.368$\pm$0.004  &43.27  &   76.7$\pm$3.9   & 62.1$\pm$3.6   & 90.4$\pm$4.7   & ...       &  9.21   &         &     & Blazar\\
\hline
Crab      & 75.355$\pm$0.010 & ...  &  0.48$\pm$0.03   & 1.25$\pm$0.07  & 0.37$\pm$0.03  & 1.27      & ...    & ....  & ...  & Pulsar \\
\enddata
\tablenotetext{1}{ASM variability strengths computed for data spanning duration of 12.5 years,
whereas BAT variability strengths correspond to data collected over 9 months. Bin size is 20
days for both the cases.}
\end{deluxetable}

The literature survey for RXTE-PCA light curves of AGNs (in this study) accumulated over 
long periods has allowed us to compare them with those of long looks of ASM. We could 
obtain PCA light curves for Mrk 501 and Mrk 421 taken over a period of 8 and 9 years 
respectively overlapping with the period for ASM light curves. The data set for Mrk 421 
covers a period 1996-2005 (Figure 1 of \citet{emm07}) and 1997-2004 for Mrk 501 \citep{gli06}.
Peaks at similar MJDs are identified on inspecting the ASM and PCA light curves 
for Mrk 421. The peaks were found near MJDs - 51000, 51700, 51900, 52600 and 53100. 
Similarly, peaks for Mrk 501 were found near the time 1997 (~MJD 50600).

\subsection{Variability strength}

We have calculated the strength of variability for all these sources in  
bin sizes of 20 and 40 days. The strength of variability and the errors 
on them are calculated as follows.

Light curve from source consists of $N$ flux measurements $x_i$ with measurement
errors of $\sigma_i$. In addition to these variations, object has intrinsic variability
or additional source variance $\sigma_Q$. It is necessary to disentangle these two
variances. One of the approaches to estimate the intrinsic variability is to use an
{\em excess variance} $\sigma_{XS}$ as an estimator \citep{nan97,vau03}. 
It is given by 
\begin{equation}
\sigma_{XS}^2 = S^2 - \overline {\sigma_i^2}
\end{equation}
where the sample variance S$^2$ is given by
\begin{equation}
S^2 = \frac{1}{N-1}\sum (x_i - \overline x)^2
\end{equation}
where $\overline x$ is mean rate
and $\overline {\sigma_i^2}$ is average variance of the measurements.

\begin{figure}
\epsscale{1.0}
\plotone{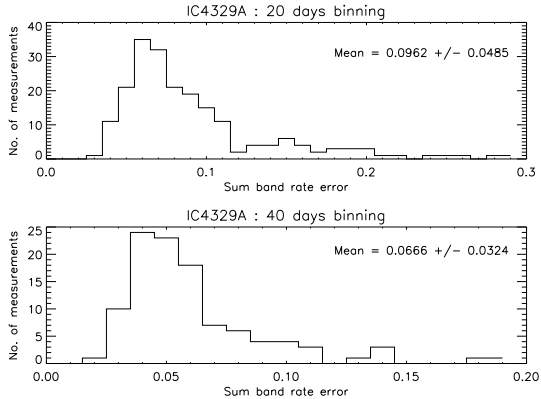}
\caption{Distribution of sum rate errors with 20 days and 40 days binning
for IC4329A after cut on number of dwells. \label{errorhist}}
\end{figure}

For light curves with varying measurement uncertainties ($\sigma_i \neq constant$), it
is necessary to use a numerical approach to obtain the best estimate for the parameter 
of interest ($\sigma_Q$, here) \citep{alm00}. The most widely used method 
for this purpose is the principle
of maximum likelihood. The probability density of obtaining data values $x_i$ is given
as a product of Gaussian functions. Using Bayes' theorem probability distribution for
$\sigma_Q$ is obtained. This is likelihood function for $\sigma_Q$ and it can be 
calculated assuming Bayesian prior distribution for $\sigma_Q$ and $x_i$. By 
differentiating, maximum likelihood estimate can be obtained for $\sigma_Q$ (see
\citet{bec07} for equations). In the case of identical measurement errors
($\sigma_i$ = constant) this expression reduces to excess variance for uniform prior.
This corresponds to $\sigma_Q$=$\sigma_{XS}$. 
In the present analysis of ASM data, since $\sigma_i$
is almost constant, we have used this simplified approach. This assumption about
approximate constancy of $\sigma_i$s is established using IC4329A data. Fig.~\ref{errorhist}
shows the distribution of sum rate error after 20 days and 40 days binning, after
applying cut on no. of dwells per bin. Mean value of binned sum rate error and
RMS is given in the figure. These distributions are sufficiently narrow to validate
the assumption of constancy of $\sigma_i$s.

Variability is given in terms of normalized excess variance, i.e.
$\sigma^2_{NXS}$=$\sigma^2_{XS}$/$\overline{x}^2$ or the fractional root mean square 
(rms) variability amplitude ($F_{var}$) given by
\begin{equation}
F_{var} = \sqrt{\frac{S^2 - \overline {\sigma_i^2}}{\overline{x}^2}}
\end{equation}
Error on variability or normalized excess variance consists of two parts: 1. arising 
from measurement error and 2. arising from intrinsic fluctuations, depending on the 
index of the power spectrum. So $err(\sigma^2_{NXS})$ is given by
\begin{equation}
[err(\sigma^2_{NXS})]^2_T = [err(\sigma^2_{NXS})]^2_M + [err(\sigma^2_{NXS})]^2_I
\end{equation}
where subscript T stands for total error, M for measurement error and I for
intrinsic error.
According to equation (11) from \citet{vau03},
\begin{equation}
[err(\sigma^2_{NXS})]^2_M=\frac{2}{N} \frac{(\overline {\sigma_i^2})^2}{\overline{x}^4} + 
\frac{4\overline {\sigma_i^2}}{N} \frac{F_{var}^2}{\overline{x}^2}
\end{equation}
Also approximating intrinsic fluctuations with white noise  (it should be
noted that the light curves are red noise and white noise is assumed here 
for simplicity) and following 
\citet{vau03},
\begin{equation}
[err(\sigma^2_{NXS})]_I = \sqrt{\frac{2}{N}} \sigma^2_{NXS} 
= \sqrt{\frac{2}{N}} F_{var}^2
\end{equation}
Combining equations 5 and 6 we get
\begin{equation}
[err(\sigma^2_{NXS})]^2_T = \frac{2}{N} \left(\frac{\overline {\sigma_i^2}}{\overline{x}^2} + F_{var}^2 \right)^2
\end{equation}
Substituting for $F_{var}$ from equation 3 we get
\begin{equation}
[err(\sigma^2_{NXS})]^2_T = \frac{2}{N} \left(\frac{S^2}{\overline{x}^2} \right)^2
\end{equation}
Hence
\begin{equation}
[err(\sigma^2_{NXS})]_T=\sqrt{\frac{2}{N}} \frac{S^2}{\overline{x}^2}
\end{equation}
Following equation (B2) of \citet{vau03},
\begin{equation}
err(F_{var})_T = \frac{1}{2F_{var}} err(\sigma^2_{NXS})
\end{equation}
This is valid when $err(\sigma^2_{NXS})$ is small and we get the following 
expression
\begin{equation}
err(F_{var})_T = \frac{1}{\sqrt{2N}} \frac{S^2}{\overline{x}^2 F_{var}}
\end{equation}
When $err(\sigma^2_{NXS})$ is large, the error estimate will only
be approximate.
The strength of variability is calculated from the expression $S_V = 100\% . 
\sigma_Q/\overline{x}$ i.e. $S_V = 100\%.F_{var}$. 

We have applied this method to downloaded light curves with different binning
(20 and 40 days), and estimated $\sigma_Q$ i.e. intrinsic
variability as well as variability strength $S_V$ for all ASM detected AGN. 
This exercise is carried out for sum band light curve over 1.5--12 keV, A
band i.e. 1.5--3 keV, as well as (B+C) band i.e. 3-12 keV. Variability
strengths for 20 and 40 days binning were found to be similar. 
Results for 20 days binning
for sum band data are given in Table~\ref{tab_var20} for 30 AGN which are common
between ASM and BAT database. 
We have also listed ASM variability strengths for four AGNs with ASM count rates
above 0.2 c/s, but not detected by BAT. Also included in the table is BAT
variability strength Mrk 3, which is not detected by ASM. In view of unknown
systematics in ASM data, we have quoted ASM variability strengths in various
bands for only sources with count rates above 0.0895 c/s for S band and 0.0274 c/s
for A band. BAT variability strengths given in this table are from \citet{bec07}.
We have quoted their results as upper limits for the sources where their 
estimates of variability strength are negative or smaller than error on estimate.

Since the X-ray spectral parameters
(like the power-law index) are not available to all these sources, we have 
calculated the  X-ray luminosities by
converting the  ASM sum band rates (R counts/s) to 
to energy flux using 
\begin{equation}
F[erg/cm^2/s]=3.2 \times 10^{-10} \times R [cts/s]
\end{equation}
This assumes Crab-like spectrum, as prescribed by  \citet{gri02}.
It should be noted that this assumption is not strictly valid
for absorbed sources. For these sources photon index, over the A band
in particular, could be much flatter than that of Crab.
For the sources where spectral parameters are available, we
get an average conversion factor very close to this value (3.0 $\times$ 10$^{-10}$ $\times$ R [cts/s]).

\begin{figure}
\epsscale{1.0}
\plotone{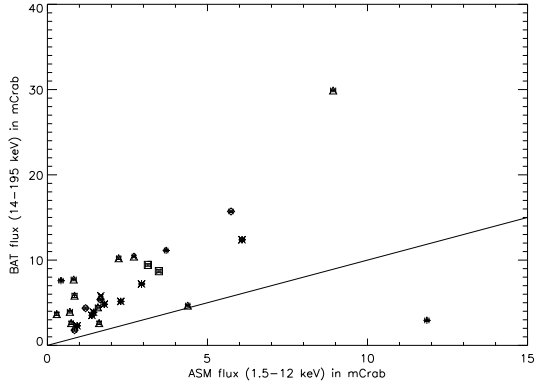}
\caption{BAT flux (14-195 keV) vs. ASM flux (1.5-12 keV) for a sample of 30 AGN
common in BAT and ASM data base. Blazars are indicated with +, Seyfert 1 by asterisk, Seyfert 1.5
by diamond, Seyfert 1.8 and 1.9 by square, Seyfert 2 by triangle and radio galaxy by cross. Solid
line indicates slope 1.
\label{asmbatflux}}
\end{figure}

To compare the flux in Crab units we have used
 a conversion factor of  75.5 counts/s (ASM sum band)
and 453.8 counts/s for BAT rates as given by \citet{bec07}. This allows
easy comparison of fluxes in two different energy bands.
Fig.~\ref{asmbatflux} shows variation of 
BAT flux vs ASM total flux in the energy range of 1.5--12 keV. Blazars are indicated 
with plus signs, Seyfert 1 with asterisks, Seyfert 1.5 with diamonds, Seyfert 1.8 and 1.9 by 
square, Seyfert 2 with triangles and radio galaxy with a  cross. Error bars are indicated for both
BAT and ASM fluxes. Typically these error bars are smaller than symbol size. 
One Blazar, Mrk 421 and one Seyfert 2 galaxy, NGC1275 (not shown in the figure)
seem to have much higher flux 
in ASM compared to other AGN with similar flux values in BAT energy band. Mrk
421 is highly variable source as can be seen from the ASM light curve. BAT
data corresponds to only 9 months out of 12.5 years of data accumulated by 
ASM. (Mrk 421 was in  a somewhat low state during the BAT observations.)  
In case of NGC 1275, ASM flux is about 33.7 mCrab whereas BAT flux is 4.5
mCrab. The field
of view of NGC 1275 contains the Perseus cluster and X-ray emission from this
cluster could be responsible for higher count rate and dilution of observed
X-ray variability as indicated by much lower variability strength for this
source \citep{san07}.
Solid line 
in the figure indicates both fluxes being linearly related (as will be the case
if the spectral slope of AGNs is crab like - a photon index of 2). BAT 
flux seems to be increasing rapidly relative to the ASM flux. 
The lack of a strong correlation between ASM and BAT fluxes
could be  due to the mixing of different types of objects: sources 
with very soft spectrum which could be preferentially observed by
the ROSAT satellite do not show flux correlation between the soft and
hard bands (Tueller, Mushotzky, Barthelmy et al. 2008), whereas
heavily obscured sources are obviously are not expected to show 
correlation (because the very soft flux in these sources may
not be originating from the central engine of the AGN).

\begin{figure}
\epsscale{1.0}
\plotone{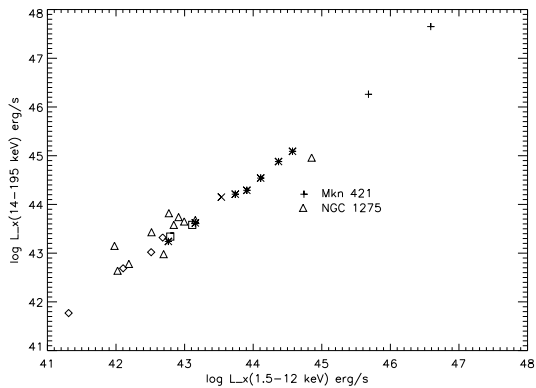}
\caption{Hard X-ray luminosity of AGN from BAT vs soft X-ray
luminosity from ASM for 30 common AGN. Blazars are indicated with +, Seyfert 1 by asterisk, Seyfert 1.5 by diamond,
Seyfert 1.8 and 1.9 by square, Seyfert 2 by triangle and radio galaxy by cross. \label{asmbatlum}}
\end{figure}

Fig.~\ref{asmbatlum} shows luminosity of AGN from BAT data as given by \citet{bec07}
vs ASM luminosity. On logarithmic scale BAT luminosity seems to be increasing with
ASM luminosity linearly. There are two AGN which are slightly away from the trend shown 
by other AGN. These are Blazar : Mrk 421 and Seyfert 2 : NGC 1275 as in the case of
Fig.~\ref{asmbatflux}. 

\begin{figure}
\epsscale{1.05}
\plotone{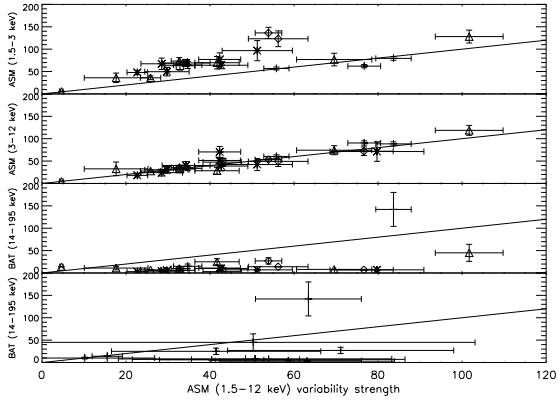}
\caption{(a) ASM variability strength in 1.5-3 keV band, (b) ASM variability strength in 3-12
keV band (c) BAT variability strength in 14-195 keV band as a function of ASM variability
strength in 1.5-12 keV band and (d) BAT variability strength in 14-195 keV band as a function of
ASM variability in 1.5-12 keV band for simultaneous data. Variability strength decreases at
higher energies.
\label{asmbatvar}}
\end{figure}

\begin{figure}
\epsscale{1.0}
\plotone{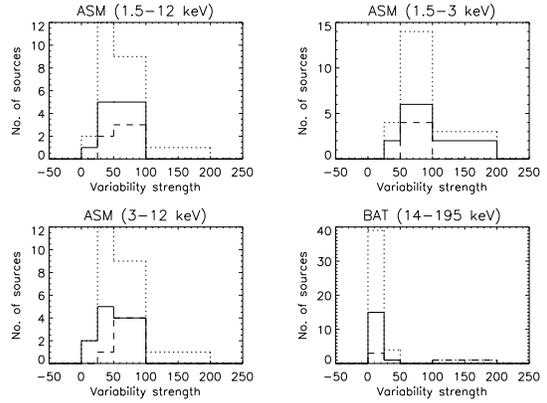}
\caption{Distribution of variability strengths of AGN in (a) 1.5--12 keV, (b)
1.5--3 keV, (c) 3--12 keV and (d) 14--195 keV as obtained from ASM and BAT
data.
In each panel, dotted line represents
distribution including all objects, dashed line corresponds to blazars and
radio galaxy and solid line represents variability strength distribution
for Seyfert 1 and 1.5.
\label{varhisto}}
\end{figure}

Fig.~\ref{asmbatvar} shows the plot of ASM and BAT variability strength in different
energy ranges. Top panel shows ASM variability strength in 1.2--3 keV band, second
panel shows ASM variability strength in 3--12 keV band and third panel shows
BAT variability strength in 14--195 keV as a function of ASM variability strength
in 1.5--12 keV range. Bottom panel also shows BAT variability strength in 14--195 keV 
as a function of ASM variability strength in 1.5--12 keV range, for near simultaneous
data. Here we have selected ASM data corresponding to first nine months of BAT
operation. Error bars on ASM variability strength in this panel are large and 
for some of the sources only upper limits on variability strength could be 
estimated for ASM data. For the sake of clarity we have restricted X-axis range
to 120 for all the panels. Bottom-most panel has three points with ASM variability
strength exceeding this value.
Solid line in each panel corresponds to slope 1. It can be
seen that variability strength decreases with increase in energy from 1.2--3 keV to
14--195 keV. Variability strength in 3--12 keV band seems to be well correlated
and almost similar to   the one in 1.5--12 keV band indicating that 
in these two energy bands most of the flux is in the large overlapping energy range. 
To investigate this further we have grouped the sources in three groups, 
1. 3 Blazars and one
radio galaxy, 2. 7 Seyfert 1 and 4 Seyfert 1.5 galaxies and 3. 1 Seyfert 1.8, 
2 Seyfert 1.9 and 11 Seyfert 2 galaxies. 
Fig.~\ref{varhisto} shows distributions
of variability strengths in 1.5--12 keV, 1.5--3 keV, 3--12 keV and 14--195 keV
bands. Dotted line corresponds histogram generated including all sources, dashed
line corresponds to blazars i.e group 1 and solid line corresponds to Seyfert 1
and 1.5, i.e. group 2. 
This figure includes BAT variability strengths given as upper limits in Table~\ref{tab_var20}.
Average values of variability strengths
for different groups and for different energy ranges are listed in Table~\ref{vartable}.
Values given here are the weighted mean and estimated error on the mean. 
BAT variability strengths given as upper limits in Table~\ref{tab_var20}
are not included here.
NGC 1275 is 
excluded here for ASM in these calculations since variability strength for this 
object quoted here could be grossly
underestimated due to contamination caused by Perseus cluster in the field of view
of ASM. This table clearly shows the trend of decrease in variability strength
with increase in energy for all types of AGN.

\begin{deluxetable}{ccccc}
\tablecolumns{5}
\tablewidth{0pc}
\tablecaption{Average variability strengths in different energy ranges \label{vartable}}
\tablehead{
\colhead{Energy Range}  & \colhead{All sources}  & \colhead{Blazars}  & \colhead{Sy 1-1.5} &\colhead{Sy 1.8-2} \\
}
\startdata
ASM (1.5-12 keV)  &   41.1$\pm$0.8   &    57.3$\pm$1.7  &   38.4$\pm$1.4   &   34.1$\pm$1.2 \\
ASM (1.5-3 keV)   &   61.4$\pm$1.5   &    64.7$\pm$2.0  &   64.0$\pm$2.9   &   51.5$\pm$3.0 \\
ASM (3-12 keV)    &   44.3$\pm$1.0   &    64.2$\pm$1.9  &   39.1$\pm$1.9   &   36.4$\pm$1.4 \\
BAT (14-195 keV)  &  12.6$\pm$1.3    &   17.2$\pm$3.6   &  19.2$\pm$5.3   &  11.4$\pm$1.5  \\
\enddata
\end{deluxetable}

Higher 
variability strength in 1.5--3 keV band compared to the one in 3--12 keV band
for Seyfert 1's is consistent with previous findings (see \citet{are08} and
references therein). They have found that the variability as a function of 
energy  peaks 
around 2 keV one time scales of one to a few days. Our result extends this property
to longer time scales. 

\begin{figure}
\epsscale{1.0}
\plotone{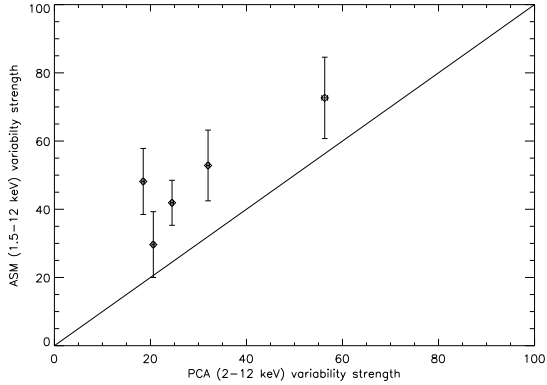}
\caption{ASM variability strength (1.5--12 keV) vs RXTE PCA variability strength (2--12 keV)
for Seyfert 1 and Seyfert 1.5 galaxies.
\label{asmpcavar}}
\end{figure}

 We have compared our results with the variability strength in 2--12 keV from
RXTE/PCA as presented by \citet{mar04}. They have given variability strength
for several Seyfert 1 and 1.5 galaxies on different time scales. There are five
sources common between their sample with timescale of 1296 days (Table 3 in their paper) 
and the present sample. These are 3C120, NGC 3783, NGC 3516, MCG-6-30-15 and NGC 3227.
For these sources we have computed average variability strength in the energy range
of 1.5 -- 12 keV with 40 days binning covering roughly the same MJD range as \citet{mar04}.
Bin size of 40 days used here is close to 34.4 days bin size used by them. 
Fig~\ref{asmpcavar} shows the comparison of ASM variability strengths in 1.5 -- 12 keV 
with PCA variability in 2 -- 12 keV range. ASM variability  seems to  be  higher
than the PCA  variability.

\subsection{Variability strength vs break time scale}

AGN variability is often expressed in terms of fluctuations in power spectral density 
(PSD) i.e. variability power $P(\nu)$ as a function of frequency, $\nu$. On longer 
timescales, the PSDs of AGN are fitted by a power-law of slope -1 which breaks to a 
steeper slope ($> 2$) at timescales shorter than the 'break' timescale, $T_B$
\citep{mar03}. Break timescale is expected to depend on the black hole mass.
Considering similarities between PSDs of X-ray binaries and AGN, break timescales
for AGN are expected to be in the range of $\lesssim$ 1 day to $\gtrsim$ a hundred days. 
Earlier attempts
of comparing the $T_B$ with black hole mass (M$_{BH}$) for AGN have shown a rough linear 
scaling but with a scatter \citep{mar03,mch04}. Improving on 
it, an inverse dependence on a second variable, probably the accretion rate, was 
suggested \citep{mch04,mch05,utt05}. 
\citet{mch06} quantified the relationship between $T_B$, M$_{BH}$ 
and $L_{bol}$ (in place of the accretion rate), as log$T_B$ = Alog$M_{BH}$ - 
Blog$L_{bol}$ + C, where M$_{BH}$ is in units of 10$^6$ solar masses, $L_{bol}$ is
bolometric luminosity in 10$^{44}$ erg s$^{-1}$. 
Best fit values determined by them are A (= 2.1), B (= 0.98) and 
C (= 2.32). Using this relation, we have calculated break timescales ($T_B$) for our
AGN sample. Values of log($M_{BH}$) (in units of $10^6 M_{\odot}$) and log($L_{bol}$) (in 
units of 10$^{44}$ erg s$^{-1}$) listed in Table~\ref{tab_var20} are adopted from 
\citet{woo02}, \citet{utt05} and \citet{wan86}. The masses
are estimated either using reverberation mapping or the BLR size-luminosity relation 
or the stellar velocity dispersion method. 
 $L_{bol}$ values are also taken from references mentioned above.
It should be noted that $L_{bol}$ for the blazars may be beamed,
leading  to a wrong estimates of $T_B$. 

\begin{figure}
\epsscale{1.0}
\plotone{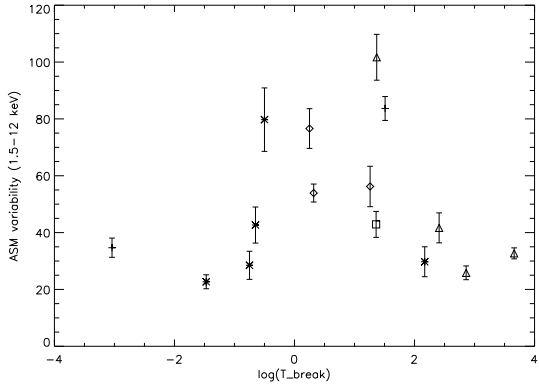}
\caption{ASM variability strength vs  break time scale $T_{break}$). \label{asmtb}}
\end{figure}

\begin{figure}
\epsscale{1.0}
\plotone{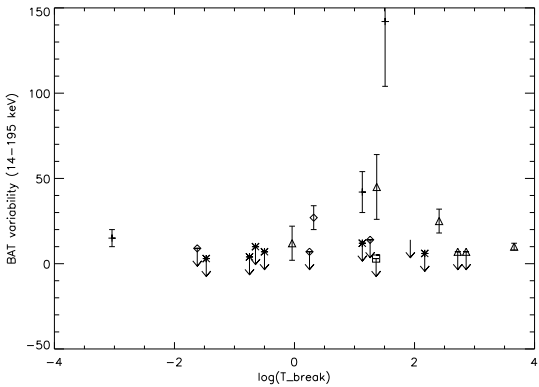}
\caption{BAT variability strength vs break time scale $T_{break}$). \label{battb}}
\end{figure}

Figure~\ref{asmtb} shows ASM variability
on the scale of 20 days as a function of log($T_B$). 
Figure~\ref{battb} shows BAT variability strength as a 
function of log($T_B$). Out of 44 AGN from BAT sample of \citet{bec07}, black
hole masses and bolometric luminosity are available for 20 sources \citep{utt05},
and hence $T_B$ could be estimated for these sources. Out of these, 19 sources are 
common between ASM and BAT database. The remaining source is Mrk 3, which is not 
in the ASM source list. 
Figure shows some
increase in variability of BAT data for AGN in the neighborhood of 20 days 
and decrease on both lower and higher sides. 
The average variability for log($T_B$) between -1 to +1.6 is 
62.9 and 53.6, for the ASM and BAT respectively and these values are
27.4 and 16.7 outside this range.
The decrease for
higher $T_B$ is as expected. The lower variability at lower $T_B$ could be
due to a variety of reasons like a) inclusion of a blazar for which
the McHardy et al (2006) relation may not be valid b) at very low value of
$T_B$, we are sampling the sources at very low frequencies where
a second turn-over of PSD is likely if the sources are a state like the
low-hard states of Galactic black hole sources.

%

\subsection{Energy spectra}

We checked for data from pointed observations of these AGN by X-Ray Telescope
(XRT) onboard Swift during this 9 month period. Data was available for four
sources from our list. These were three Blazars : 3C273, 3C454.3 and Mrk 421 and
one Seyfert 2 Galaxy : Cen A. We have fitted spectrum over the energy range of 
0.3--10 keV for these four objects. For each of these objects source and background 
photons were selected using XSELECT version 2.4. For Photon Counting (PC) mode source 
photons were selected in a circular region with the radius of 20 pixels (i.e ~ 47 
arcseconds), whereas background photons were extracted from nearby circular region 
with a radius of 40 pixels. For data collected in Windowed Timing (WT) mode, source 
photons were extracted using box region with the length of 40 pixels and width about 
20 pixels. Events with grades 0-12 and 0-2 were selected for PC and WT mode data, 
respectively. The spectral data were rebinned by GRPPHA 3.0.0 with 20 photons per 
bin. Standard auxiliary response files and response matrices were used.

Spectra for these sources were fitted using XSPEC version 12.3.1 with a model
consisting of absorbed power law over the energy range of 0.3--10 keV. In each case 
$N_H$ was fixed to the value given in Table 2 of \citet{bec07}. Power law indices 
and 2--10 keV flux obtained from fit are given in Table~\ref{tab_spec} . 
It should be noted here, however, that 
the Swift 
observations may not be equal to the average photon index
representing the ASM observations.

Among the three Blazars, Mrk 421 has the steepest spectrum and 3C 454.3 has the
flattest one. In case of 3C454.3, we have not estimated the variability strength in 
ASM band because of low count rate. ASM variability strengths for other two
blazars are : 83.7$\pm$4.2 for Mrk 421 and 34.7$\pm$3.4 for 3C273. Whereas BAT
variability strengths of these three sources arranged in the order of increasing 
power-law index are : 3C454.4 (42$\pm$12), 3C273 (15$\pm$5) and Mrk 421 (142$\pm$38).
This indicates an increase in variability strength for sources with steeper spectra.
But this inference should be taken with caution. It should be noted that in case
of Mrk 421, which is a BL Lac type of object, the hard X-ray band is dominated by high
energy synchrotron emission whereas in case of 3C454.3 and 3C273, which are FSRQ,
hard X-ray emission is mainly inverse-Compton emission. This could be the one possible 
reason for steeper spectrum of Mrk 421 compared to the other two objects. 
It is also quite possible that the long duration variabilities of blazars are 
influenced by infrequent strong flare events and the higher variability in any
one blazar could simply be the effect of a few large events which happen to occur
in that source.


\begin{deluxetable}{ccccc}
\tablecolumns{5}
\tablewidth{0pc}
\tablecaption{Spectral Fits \label{tab_spec}}
\tablehead{
\colhead{Source}  &  \colhead{Observation}  & \colhead{$N_H$} & \colhead{Powerlaw}  &
\colhead{2-10 keV} \\
\colhead{name} & \colhead{duration} & \colhead{10$^{22}$cm$^{-2}$} & \colhead{index}  &
\colhead{flux} \\
\colhead{} & \colhead{s}  & \colhead{} & \colhead{} &  \colhead{erg/cm$^2$/s}
}
\startdata
3C 273   & 2962.6 & 0.0316 & 1.79$\pm$0.030 & 5.87$\times$10$^{-11}$ \\
3C 454.3 & 13670.0 & 0.0631 & 1.32$\pm$0.02 & 5.43$\times$10$^{-11}$\\
Mrk 421  & 6958.0 & 0.001 & 2.38$\pm$0.01 & 8.34$\times$10$^{-11}$\\
Cen A    & 7699.5 & 12.59 & 1.64$\pm$0.06 & 1.99$\times$10$^{-10}$\\
\enddata
\end{deluxetable}

\section{Discussion and conclusions}

Study of X-ray variability of AGN at longer time scales and also at different
energies is very important to pin down the accretion disk geometry and
the radiation processes involved in the X-ray emission very close to the 
black holes. We have made a systematic study of  the soft X-ray variability
characteristics of all AGN with measured long time
scale variability from Swift/BAT. 

 One of the important findings of this work is that individual ASM dwells
can be co-added to obtain flux integrated over long time scales and
by propagating the measurement errors, very
low errors on the data points are obtained. This method assumes that most
of the systematic errors in the flux measurements in the individual 
dwell measurements are understood and taken care of. Support for
this assumption comes from the fact that the long duration light curves
of several AGN bear a striking similarity to the light curves
obtained from pointed RXTE/PCA observations. Some more work, however,
need to be done to understand possible time dependent systematic
errors so that a complete data set can be used to derive meaningful
power spectral densities.




  Another important finding is the quite uniform variation of
variability strength with energy for diverse classes of objects,
except for the blazar Mrk 421. For this particular source, there is
a marked increase in the variability as a function of energy. Such
behavior has been observed at shorter durations.
\citet{hor09} made a detailed multi-wavelength study
for a period of about 300 days and   find the variability to increase
across the full electro-magnetic band. In particular, they
noticed a sharp increase of variability from soft X-rays
to hard X-rays: the values for variability (F$_{var}$) 
were  26.9, 44.3, 52.9 and 99.3,
respectively for Swift XRT, RXTE-ASM, RXTE-PCA and Swift-BAT data
(for observation bins of a few tens of thousand seconds and 
duration
in the range of 24 to 256 days).
The values that we obtained are 79 (RXTE-ASM 1.5 -- 3 keV),
88 (RXTE-ASM 3 -- 12 keV) for 11 years and 142 (Swift-BAT for
300 days). The energy dependent variability behavior of the the blazar
in our list, 3C 273,  for which data is available both in the soft and hard
X-rays, shows an energy dependent behavior similar to Seyfert 1
galaxies. \citet{mch06a} has noted that the wide band PSD
of 3C273 is identical to that  of Seyfert galaxies and contended that
the process responsible for variability in this source is the same that
produces variability in non-beamed sources.

Most of the other sources in our sample are Seyfert galaxies and a 
decrease in the variability as a function of energy is observed.
In a detailed study of a large number of Seyfert galaxies based on
the first seven years of RXTE-PCA monitoring, \citet{mar04} 
give energy dependent variability data for long durations and 
a similar trend of decrease in variability with energy has been noticed.
There are five Seyfert 1 galaxies (3C120, 3C390.3, IC4329A, NGC3783,
and MCG-6-30-15) and  two Seyfert 1.5 galaxies (NGC4151 and NGC3227)
common between their sample and our work. The ratio of variability
for the soft (2 -- 4 keV) and the hard (7 -- 12 keV) band are,
respectively, 1.15, 1.26, 1.19, 1.28, 1.37, 1.26 and 1.35, for a
duration of 216 days. For
comparison similar ratios from our data
(1.5 -- 3 keV vs 3 -- 12 keV), for the above sources, respectively, are,
1.65, 1.62, 2.65, 2.9, 1.71, 2.6, and 2.5, for the full ASM
duration.  The typical
uncertainties are 0.06 for the RXTE data and 0.3 for our
data.  Though this ratio from RXTE data is fairly same for all the
sources (about 1.27), for the ASM data it varies from 1.66 to 2.64.
For five sources energy dependent variability at
longer times scales ($>$ 16 days) is given by \citet{mar03}
and for two of the sources where we have
obtained very high ratio (NGC 3783 and NGC 4151) a sharp upturn
towards lower energies are noticed (see their Fig 6). 
Further, RXTE-ASM is more sensitive
to low energy X-rays. The effective area at 2 keV is about 33\% of  the
effective area at 5 keV \citep{lev96}, whereas this number is $<$ 1\%
for RXTE PCA \citep{jah06}, mainly due to the propane layer
on the top side of RXTE-PCA  which has lower transparency at 
low energies due to aluminized mylar windows and the permeation of xenon
gas into the propane layer \citep{jah96}. 
Hence it is quite possible that the higher variability
at low energies derived by us is real and indicate a sharp
increase of variability
in AGNs at lower energies for longer time scales.
Though we cannot completely rule out some further systematic errors,
it is quite evident that RXTE-ASM is an unique instrument to probe
long time scale variability at lower energies.

There are, indeed, several indicators from early EXOSAT data
which supports the above assertion. \citet{wal92}
discuss X-ray variability of several AGNs observed by EXOSAT over
the period of an year and note that soft X-rays vary by factors up to
7, much higher than that seen in hard X-rays. Two of the sources
in their list is common to the present work: MR2251-178 and NGC 3783.
For these two sources, they report a peak-to-peak variation of
4.5 and 3, respectively  in the
soft X-rays, compared to 1.9 and 1.8 in hard X-rays.
This should be compared with the $S_V$ derived by us for these two
sources: 97 and 67, respectively in the
1.5 -- 3 keV band
and 42$\pm$14 and 24$\pm$7 in the 3 -- 12 keV band, respectively.
Though there is a strong indication of very high soft X-ray
variability at long time scales, a detailed comparison with
a large number of pointed observations for individual sources are
required to pin down any residual systematic errors in the RXTE-ASM
data.

Energy dependent variability has been studied extensively 
in Galactic black hole sources (see \citet{zdz05} and references
therein) and one of the models used to explain
the energy dependency is the variations in the input parameters
of the thermal Comptonization process which results in an
pivoting power-law spectrum.
Though the very low value of variability in the BAT energy band
may be influenced by a non-varying spectral component in this
energy band (like reflection), detailed wide band spectroscopy
of Seyfert galaxies suggest a common continuous phenomena
for the decreasing variability with increasing energy.
For example, the radio galaxy 3C120 shows only a moderate
reflection component and very strong spectral variability (over
two days) with the spectral index correlated with the soft X-ray
flux, indicating a phenomena where the input seed photon
variation for a thermal Comptonization process causes the 
energy dependent X-ray variability \citep{zdz01}.
If the same phenomena is responsible for most of the sources
reported in this work,
it suggests that a)
for a majority of AGNs the thermal Comptonization is
the dominant process, b) the pivot energy is higher (greater than about
20 keV). \\

This work is based  on 
results provided by the ASM/RXTE teams at MIT and at the 
RXTE SOF and GOF at NASA's GSFC. This research has also 
 made use of data obtained from the High Energy Astrophysics Science 
Archive Research Center (HEASARC), provided by NASA's Goddard Space Flight Center.
We are very grateful to 
the referee of this paper for the very thoughtful and critical comments.

\end{document}